\documentclass[sigconf,natbib=true,anonymous=False]{acmart}
\usepackage{multirow}
\usepackage{colortbl}

\AtBeginDocument{%
  }

\copyrightyear{2025}
\acmYear{2025}
\setcopyright{rightsretained}\acmConference[COLIEE 2025]{COLIEE 2025 workshop: Competition on Legal Information Extraction/Entailment}{June 20, 2025}{Chicago, USA}
\acmBooktitle{Proceedings of COLIEE 2025 workshop, June 20, 2025, Chicago, USA}
\acmPrice{}
\acmISBN{}
\acmDOI{}

\begin{document}

\title{UQLegalAI@COLIEE2025: Advancing Legal Case Retrieval with Large Language Models and Graph Neural Networks}

\author{Yanran Tang}
\email{yanran.tang@uq.edu.au}
\affiliation{\institution{The University of Queensland}
  \city{Brisbane}
  \country{Australia}}
\author{Ruihong Qiu}
\email{r.qiu@uq.edu.au}
\affiliation{\institution{The University of Queensland}
  \city{Brisbane}
  \country{Australia}}
\author{Zi Huang}
\email{helen.huang@uq.edu.au}
\affiliation{\institution{The University of Queensland}
  \city{Brisbane}
  \country{Australia}}

\renewcommand{\shortauthors}{Tang et al.}

\begin{abstract}
Legal case retrieval plays a pivotal role in the legal domain by facilitating the efficient identification of relevant cases, supporting legal professionals and researchers to propose legal arguments and make informed decision-making. To improve retrieval accuracy, the Competition on Legal Information Extraction and Entailment (COLIEE) is held annually, offering updated benchmark datasets for evaluation. This paper presents a detailed description of CaseLink, the method employed by UQLegalAI, the second highest team in Task 1 of COLIEE 2025. The CaseLink model utilises inductive graph learning and Global Case Graphs to capture the intrinsic case connectivity to improve the accuracy of legal case retrieval. Specifically, a large language model specialized in text embedding is employed to transform legal texts into embeddings, which serve as the feature representations of the nodes in the constructed case graph. A new contrastive objective, incorporating a regularization on the degree of case nodes, is proposed to leverage the information within the case reference relationship for model optimization. The main codebase used in our method is based on an open-sourced repo of CaseLink~\cite{caselink}:  ~\url{https://github.com/yanran-tang/CaseLink}.
\end{abstract}





\maketitle

\section{Introduction}
In legal domain, a precedent refers to a judicial decision that serve as an example or authority when giving judgment to future similar cases, which ensures fairness, consistency and predictability of judgments in legal system. However, identifying relevant precedents within large legal databases is a time-consuming task, significantly reducing the work efficiency of legal practitioners. Therefore, the Competition on Legal Information Extraction and Entailment (COLIEE)~\cite{COLIEE} is held annually to encourage the competition participants to develop highly accurate legal case retrieval models. There are five task in COLIEE 2025, where Task 1 and Task 2 are case law tasks, Task 3 and Task 4 are statute law tasks and Task 5 is a pilot task that focuses on judgment of civil cases.

In COLIEE 2025, Task 1 is a legal case retrieval task of case law system, aimed at returning `noticed cases' from a large case collection for a given query case. Specifically, a case called `noticed' to a query case means the case is referenced by the query case. The provided cases of Task 1 are all from the Federal Court of Canada. As legal case retrieval is a basic and essential task in COLIEE, previous teams have proposed high-accuracy retrieval models. In COLIEE 2024, TQM~\cite{TQM} team achieves the first place in Task 1 by exploring various lexical and semantic retrieval models. THUIR~\cite{THUIR@COLIEE} team develops a structure-aware pre-trained language model called SAILER to improve the model understanding ability of legal cases, which ranks the first in COLIEE 2023. While UA~\cite{UA@COLIEE2022} team leverages a transformer-based model for generating paragraph embeddings and a gradient boosting classifier to decide a case is noticed or not, which ranks the first in COLIEE 2022.

In this paper, our novel CaseLink~\cite{caselink} model utilised in COLIEE 2025 Task 1 is proposed to further enhance the retrieval accuracy by leveraging case connectivity relations and graph-based model. Firstly, the training set and test set are transferred into two Global Case Graphs (GCG) by exploiting the Case-Case and Case-Charge and Charge-Charge relationships of cases and charges. With the constructed graph, a large language model specialized in text embedding is utilized to convert legal texts into embeddings as the node features of GCG. To leverage the connected relationships in GCG, a graph neural network module is used to generate the case representation. A contrastive loss and a degree regularisation are designed to train the CaseLink model. Our team, UQLegalAI, ranked as the second highest team in Task 1 with a F1 score of 0.2962.

\section{Related Work}
Legal case retrieval aims to retrieve a set of relevant cases for a query case within a large legal case dataset. The recent legal case retrieval can be roughly divided into lexical models, semantic models and graph-based models. The traditional lexical models like TF-IDF~\cite{TF-IDF}, BM25~\cite{BM25} and LMIR~\cite{LMIR} utilise term frequency to calculate the case similarity score. While the semantic models such as BERT-PLI~\cite{BERT-PLI} and PromptCase~\cite{promptcase} are both using language model to generate case embedding. 

Unlike lexical and semantic models, graph-based models leverage graph structures and Graph Neural Networks (GNNs) to enhance the performance of legal case retrieval. For example, CaseGNN\cite{casegnn} and CaseGNN++\cite{casegnn++} exploit the relations of legal elements of a case to construct a fact graph and a legal issue graph for each case as well as use GNN to generate case graph representations. SLR \cite{SLR} and CFGL-LCR \cite{cfgl} integrate external knowledge graphs with GNNs to improve retrieval performance. In contrast to these graph-based models, the CaseLink~\cite{caselink} model utilised in this paper fully exploits the connectivity relationships among cases of a large legal dataset to enhance case retrieval effectiveness.

\section{Preliminary}
\subsection{Task Definition}
Task 1 of COLIEE 2025 is a legal case retrieval task that focuses on case law. Given a query case $q\in\mathcal{D}$, and a set of $n$ cases $\mathcal{D}=\{d_1,d_2,...,d_n\}$, the task of legal case retrieval is to extract a set of relevant cases $\mathcal{D}^* = \{d^*_i| d^*_i \in \mathcal{D} \wedge relevant (d^*_i, q) \}$. The $relevant (d^*_i, q)$ indicates that $d^*_i$ is a relevant case to the query case $q$. In case law, the above relevant cases refer to the precedents, which are the prior cases referenced by the query case.

\subsection{Dataset}
\begin{table}[!t]\centering
\caption{Statistics of Task 1 dataset.}\label{tab:dataset}
\small
    \begin{tabular}{c|cc}
    \toprule
    COLIEE 2025 Task 1 &train &test\\
    \cmidrule{1-3}
    Language &\multicolumn{2}{c}{English} \\
    \# Query &1678 &400 \\
    \# Candidates &7350 &2159 \\
    \# Avg. relevant cases &4.1007 &4.3975 \\
    Avg. length (\# token) &28865 &31250 \\
    Largest length (\# token) &650534 &681027 \\
    \bottomrule
    \end{tabular}
\end{table}

The cases in the Task 1 dataset are sourced entirely from the Federal Court of Canada. The statistics for the Task 1 dataset are presented in Table~\ref{tab:dataset}. These statistics reveal that the number of queries and candidates in the training set are approximately three times greater than that in the test set. Additionally, the average number of relevant cases per query in both the training and test sets are close to four, suggesting that the level of difficulty for both sets is comparable. Furthermore, the average token count per case for both the training and test sets is approximately 30,000. Notably, the longest case contains up to 680,000 tokens, highlighting the challenges associated with processing and comprehending long cases.

\subsection{Evaluation metric}
In COLIEE 2025 Task 1, the micro-average of precision, recall, and F-measure are utilised as the evaluation metic as follows:
\begin{equation}
\label{eq:precision}
\text{Precision}=\frac{\text{the number of correctly retrieved cases for all queries}}{\text{the number of retrieved cases for all queries}},
\end{equation}

\begin{equation}
\label{eq:recall}
\text{Recall}=\frac{\text{the number of correctly retrieved cases for all queries}}{\text{the number of relevant cases for all queries}},
\end{equation}

\begin{equation}
\label{eq:f1}
\text{F-measure}=\frac{2 \times \text{Precision} \times \text{Recall}}{\text{Precision} + \text{Recall}}.
\end{equation}

\section{Method}

\subsection{Global Case Graph}
\begin{figure}[!t]
\centering
\includegraphics[width=0.8\linewidth]{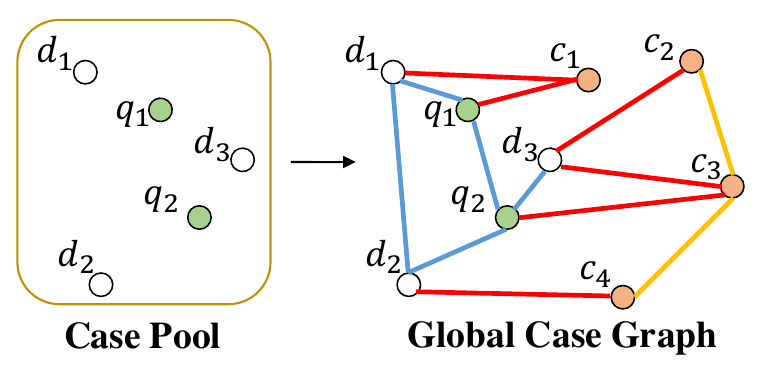}
\caption{An example of a Global Case Graph is shown, where green nodes represent the two query cases $q_1$ and $q_2$, white nodes denote candidate cases $d_1\sim d_3$ and orange nodes correspond to legal charges $c_1\sim c_4$. The solid lines indicate the edges: Case-Case edges are shown in blue, Case-Charge edges in red, and Charge-Charge edges in yellow.}
\label{fig:gcg}
\end{figure}

In this paper, the construction method of the Global Case Graph (GCG) is adopted from our previous work CaseLink~\cite{caselink}. Specifically, GCG is represented as $G = (\mathcal{V}, \mathcal{E})$, where $\mathcal{V}$ and $\mathcal{E}$ denote the set of nodes and the set of edges, respectively. The node set $\mathcal{V}$ comprises both case nodes $d, q$ as well as charge nodes $c$. The edge set $\mathcal{E}$ encompasses three types of edges: Case-Case edges, Case-Charge edges, and Charge-Charge edges. An example of GCG is shown in Figure~\ref{fig:gcg}.

\subsubsection{Nodes}
In GCG, the case nodes refer to the cases in the $\mathcal{D}$, which includes both the query cases $q$ and candidate cases $d$. The charge nodes are derived from a list specified in the Federal Courts Act and Rules of Canada\footnote{\url{https://www.fct-cf.gc.ca/en/pages/law-and-practice/acts-and-rules/federal-court/}}, presented as $\mathcal{C}=\{c_1,c_2,...,c_m\}$.

\subsubsection{Edges}
To effectively utilise intrinsic case connectivity relations, three edge connection strategies are exploited in GCG:

\noindent$\bullet$ Case-Case Edge. The construction of GCG aims to establish edges between cases that employ intrinsic connectivity. Therefore, the high similarity cases measured by BM25~\cite{BM25} will be linked as neighbour nodes. The adjacency matrix of Case-Case edges $\mathbf{A}_{d} \in \mathbb{R}^{n\times n}$ is denoted as:
    \begin{equation}
        \label{eq:d-edge}        \mathbf{A}_{d_{ij}}=\left\{
        \begin{aligned}
        1 \quad & \text{for} \text{TopK}(\text{BM25}(d_{i},d_{j}|d_{i},d_{j} \in \mathcal{D})), \\
        0 \quad & \text{for} \text{Others}, 
        \end{aligned}
        \right.
    \end{equation}
where $d_{i}$ and $d_{j}$ are two cases in $\mathcal{D}$. TopK retrieves the top $K$ most similar cases from a given list of BM25 case similarity scores.

\noindent$\bullet$ Charge-Charge Edge. In legal system, natural relationships exist among different legal charges. Therefore, connecting similar charges can effectively enhance case representation learning. The Charge-Charge edges symmetric adjacency matrix $\mathbf{A}_{c}\in\mathbb{R}^{m\times m}$, comprising $m$ charges is defined as:
\begin{equation}
        \label{eq:c-edge}
        \mathbf{A}_{c_{ij}}=\left\{
        \begin{aligned}
        1 \quad & \text{for} & \text{Sim}(\textbf{x}_{c_{i}},\textbf{x}_{c_{j}}|c_{i},c_{j} \in \mathcal{V})>\delta, \\
        0 \quad & \text{for} & \text{Others}, 
        \end{aligned}
        \right.
\end{equation}
where Sim is the similairty calculation fuction such as cosine similarity, $c_{i}\in\mathcal{V}$, $c_{j}\in\mathcal{V}$ are two charge nodes with the node features $\textbf{x}_{c_{i}} \in \mathbb{R}^d$, $\textbf{x}_{c_{j}} \in \mathbb{R}^d$. The number of Charge-Charge edges is regulated by a similarity score threshold $\delta$.

\noindent$\bullet$ Case-Charge Edge. A Case-Charge edge is established when a charge name appears in the case, which shows the high correlation between the charge and case. The adjacency matrix of Case-Charge edges $\mathbf{A}_{b}\in\mathbb{R}^{m\times n}$ is designed as:
\begin{equation}
\label{eq:cd-edge}
\textbf{A}_{b_{ij}}=\left\{
\begin{aligned}
1 \quad & \text{for} & t_{c_{i}} \ \text{appears in} \ t_{d_{j}}, \\
0 \quad & \text{for} & \text{Others},
\end{aligned}
\right.
\end{equation}
where $t_{c_{i}}$ is the text of charge $i$, $t_{d_{j}}$ is the text of case $j$.

\noindent$\bullet$ Overall Adjacency Matrix. To directly combine the Case-Case edges, Case-Charge edges, and Charge-Charge edges, the GCG overall adjacency matrix $\mathbf{A} \in \mathbb{R}^{(n+m)\times (n+m)}$ is undirected and unweighted, which is denoted as:
\begin{equation}
 \mathbf{A} = \begin{bmatrix}
 \mathbf{A}_{d}  & \mathbf{A}_{b}^\intercal \\
 \mathbf{A}_{b} & \mathbf{A}_{c} 
 \end{bmatrix},
\end{equation}
where $\textbf{A}_{b}^\intercal$ denotes the transpose of the adjacency matrix.

\subsubsection{Embedding Initialisation with Large Language Models}
Given the high quality of text embedding encoded by large language models in recent text embedding benchmark such as Multilingual Text Embedding Benchmark (MTEB)~\cite{mteb}, LLM is employed to encode the nodes into embedding features for GCG in this paper. The encoding process is denoted as:
\begin{equation}
\label{eq:casenode}
    \mathbf{x} = \text{LLM}(t),
\end{equation}
where $t$ is the text of a case or a charge, $\textbf{x} \in \mathbb{R}^d$ is the generated text embedding as the node feature in GCG. $\text{LLM}$ can be any LLM that encodes the texts into embeddings. In this paper, the top-ranked open-source model for legal retrieval task in MTEB, e5-mistral-7b-instruct~\cite{mistral}, is chosen to be the LLM text encoder.

\subsection{CaseLink}
\begin{figure}[!t]
\centering
\includegraphics[width=1\linewidth]{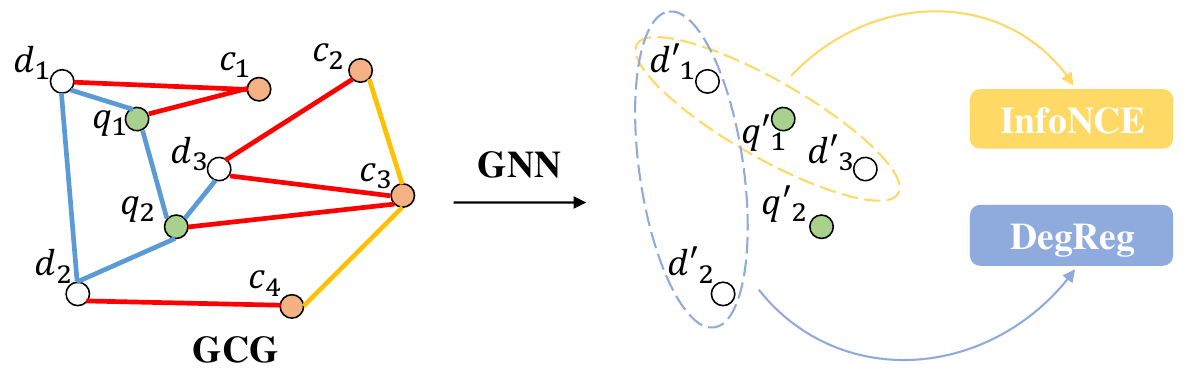}
\caption{The overall framework of CaseLink~\cite{caselink}.}
\label{fig:caselink}
\end{figure}

The CaseLink module is adopted from our previous work~\cite{caselink}, which demonstrated strong performance on the legal case retrieval task. During training, the training queries, candidate cases, and legal charges are integrated into a GCG. A graph neural network (GNN) module is then applied to update the node features within the GCG. The updated features of the query and candidate nodes are subsequently used in two training objectives: contrastive learning via the InfoNCE loss and the degree regularization (DegReg) objective, to optimize the CaseLink model. The overall framework of CaseLink is demonstrated in Figure~\ref{fig:caselink}.

\subsubsection{Graph Neural Network}
With the constructed GCG and the encoded initial graph features, a GNN model is leveraged to generate the case representation as:
\begin{equation}
\label{eq:gnnmodel}
    \textbf{H}=\text{GNN}_{\theta}(\textbf{X},\mathbf{A}),
\end{equation}
where $\mathbf{X}$ is the feature matrix consists of the node features $\mathbf{x}$, $\mathbf{H} \in \mathbb{R}^{n\times d}$ are the representations of cases in GCG, $\mathbf{h}_{i}\in\mathbb{R}^d$ is the case representation of case $i$ and $\theta$ is the model parameter of GNN. $\text{GNN}_{\theta}$ can be any graph neural network models, such as GCN~\cite{GCN}, GAT~\cite{GAT} or GraphSAGE~\cite{GraphSAGE}. 

\subsubsection{InfoNCE Objective.}
A widely adopted approach for training the GNN model in legal case retrieval is to utilize contrastive learning based on the InfoNCE objective~\cite{infonce}:
\begin{equation}
\label{eq:infonce}
\ell_{\text{InfoNCE}}=-\text{log}\frac{e^{\text{Sim}(\mathbf{h}_{q},\mathbf{h}_{d^+})/\tau}}{e^{\text{Sim}(\mathbf{h}_{q},\mathbf{h}_{d^+})/\tau}+\sum^{p}_{i=1}e^{\text{Sim}(\mathbf{h}_{q},\mathbf{h}_{d^{-}_i})/\tau}},
\end{equation}
where a relevant case $d^+$ and $p$ irrelevant cases $d^-$ are sampled for a given query case $q$ with $\tau$ denoting the temperature parameter. Cosine similarity is chosen as the Sim function here. The positive samples correspond to the ground-truth labels, and the easy negative samples are randomly sampled from training set. Specifically, the hard negative samples are randomly selected from the negative cases of Top-K BM25~\cite{BM25} ranking list.

\subsubsection{Degree Regularisation.} 
To implement degree regularisation for the candidate nodes, the pseudo-adjacency matrix is defined based on the updated node features after GNN computation as:
\begin{equation}
    \hat{\mathbf{A}}_{ij} = \text{cos}(\mathbf{h}_{i},\mathbf{h}_{j}),
\end{equation}
where $\mathbf{h}_{i}$ and $\mathbf{h}_{j}$ are the updated features of case node $i$ and $j$ in the case pool $\mathcal{D}$. The matrix $\hat{\mathbf{A}}\in\mathbb{R}^{n\times n}$ indicates a fully connected situation. And the degree regularisation is conducted on this $\hat{\mathbf{A}}$ only for candidate cases:
\begin{equation}
\label{eq:degreg}
    \ell_{\text{DegReg}} = \sum\limits^o_{i=1}\sum\limits^{n}_{j=1}(\hat{\mathbf{A}}_{ij}),         
\end{equation}
where $o$ is candidate number. 

\subsubsection{Overall Objective.}
The overall objective is designed as:
\begin{equation}
\label{eq:loss-overall}
    \ell = \ell_{\text{InfoNCE}}+\lambda\cdot\ell_{\text{DegReg}},
\end{equation}
where $\lambda$ is the coefficient of degree regularisation.

\subsection{Inference}
During testing on $\mathcal{D}_{\text{test}}$, the similarity score $s_{(q,d)}$ is calculated as:
\begin{equation} 
\label{eq:inference}
    s_{(q,d)} = \text{Sim} (\mathbf{h}_{q}, \mathbf{h}_{d}),   
\end{equation}
where $\mathbf{h}_{q}$ and $\mathbf{h}_{d}$ are the representations of query $q$ and candidate $d$ generated by CaseLink. Final top 5 ranking candidates are retrieved.

\subsection{Post-processing}
After the calculation of similarity score, the following post-processing strategies are conducted for improving retrieval accuracy.

\subsubsection{Two-stage Ranking}
To harness the strengths of statistical methods, the candidate lists are initially reduced to ten cases using the BM25 retrieval algorithm. Considering that the average number of relevant cases in the training set is 4.1, the number of final retrieved cases per query is fixed at five during the testing phase, based on the calculated CaseLink similarity scores.

\subsubsection{Year Filtering}
It is known that precedents is the prior judicial decision that serves as an example for future cases, which means that the cited precedents should happen before the given query case. Therefore, given a query, only cases with earlier dates than the query case are considered as candidates, while those with later dates are excluded in this paper. Specifically, the latest date appearing in a case is taken as its representative trial date.

\section{Experiments and Results}

\subsection{Implementation}
The training batch size is selected from \{256, 512, 1024, 1678\}. The default GNN model is GAT~\cite{GAT} with number of layers chosen from \{1,2,3\}. The dropout~\cite{dropout} rate is selected from \{0.1, 0.2, 0.5\}. The default optimiser is Adam~\cite{Adam} with learning rate chosen from \{1e-2, 1e-3, 1e-4\} and the weight decay values from \{1e-3, 1e-4, 1e-5\}. In contrastive training, each query is associated with one positive sample and one easy negative sample, while the number of hard negative samples is selected from \{1, 5, 10\}. In-batch samples of other queries are also treated as easy negative samples. For degree regularisation, the coefficient $\lambda$ is chosen from \{0,5e-4,1e-3,5e-3\}. The number of TopK case neighbour node $K$ in Equation~\ref{eq:d-edge} is selected from \{3, 5, 10\}. The threshold $\delta$ in Equation~\ref{eq:c-edge} is chosen from \{0.85, 0.9, 0.95\}. Due to the 4096 token limit of e5-mistral-7b-instruct model, any case exceeding this length is truncated to 4096 tokens.

\subsection{Result}

\begin{table}[!t]\centering
\caption{Top 10 runs of Task 1.}\label{tab:result}
\resizebox{1\linewidth}{!}
{
\begin{tabular}{ccccc}
\toprule
\textbf{Team} &\textbf{Submission} &\textbf{Precision} &\textbf{Recall} &\textbf{F1}\\
\midrule
JNLP &jnlpr\&fe2.txt &0.3042 &0.3735 &0.3353 \\
JNLP &jnlpr\&fe1.txt &0.2945 &0.3667 &0.3267 \\
\rowcolor{lightgray}
UQLegalAI &uqlegalair3.txt &0.2908 &0.3019 &0.2962\\
\rowcolor{lightgray}
UQLegalAI &uqlegalair2.txt &0.2903 &0.3013 &0.2957\\
\rowcolor{lightgray}
UQLegalAI &uqlegalair1.txt &0.2886 &0.2996 &0.2940\\
AIIR Lab &task1.aiirmpmist5.txt &0.2040	&0.2319 &0.2171\\
NOWJ &
    \begin{tabular}{@{}c@{}}prerank\_dense\_bge-rerank\_ \\ bge\_ft\_llm2vec\_major\_vote.txt\end{tabular}
    &0.1670 &0.2445 &0.1984\\
AIIR Lab &task1.aiircombmnz.txt	&0.2317	&0.1580 &0.1879 \\
AIIR Lab &task1.aiirmpmist3.txt	&0.2308	&0.1575 &0.1872 \\
NOWJ &prerank\_dense\_bge-rerank\_bge\_ft.txt &0.1605	&0.1825 &0.1708\\
\bottomrule
\end{tabular}
}
\end{table}

The final top 10 runs of COLIEE 2025 Task 1 is shown in Table~\ref{tab:result}. The three runs of Team UQLegalAI are all run by CaseLink model with different hyperparameters. The F1 score reaches 0.2962, with a gap of less than 0.04 compared to the first-place team.

In addition to the overall performance, our method exhibits a \textbf{stable} situation. The result is stable that the variance of Precision, Recall and F1 is small. The gap is less than $\pm0.003$ for our method. While for other methods, such as JNLP's submission, the variance is around $\pm0.01$.

Compared with the highest performance, our method has a lower Recall score. This can be due to the selection of more candidates in ranking.

\section{Conclusion}
This paper presents the approach of Team UQLegalAI for Task 1 of the COLIEE 2025 competition. To leverage the intrinsic connectivity relationships between legal cases, a method called CaseLink is leveraged. Within the CaseLink framework, a Global Case Graph construction module is introduced to build a case graph comprising case-case edges, case-charge edges, and charge-charge edges for each case. Node features within the GCG are encoded by a high-quality text embedding large language model. A graph neural network module is then employed to generate informative case representations. The CaseLink model is trained with an InfoNCE contrastive objective combined with a novel degree regularization term. The final ranking results for Task 1 demonstrate the effectiveness and strong performance of the proposed CaseLink approach. In future work, developing more powerful models is still needed for enhancing the accuracy of legal case retrieval.


\bibliographystyle{ACM-Reference-Format}
\bibliography{sample-base}


\begin{thebibliography}{22}


\ifx \showCODEN    \undefined \def \showCODEN     #1{\unskip}     \fi
\ifx \showISBNx    \undefined \def \showISBNx     #1{\unskip}     \fi
\ifx \showISBNxiii \undefined \def \showISBNxiii  #1{\unskip}     \fi
\ifx \showISSN     \undefined \def \showISSN      #1{\unskip}     \fi
\ifx \showLCCN     \undefined \def \showLCCN      #1{\unskip}     \fi
\ifx \shownote     \undefined \def \shownote      #1{#1}          \fi
\ifx \showarticletitle \undefined \def \showarticletitle #1{#1}   \fi
\ifx \showURL      \undefined \def \showURL       {\relax}        \fi
\providecommand\bibfield[2]{#2}
\providecommand\bibinfo[2]{#2}
\providecommand\natexlab[1]{#1}
\providecommand\showeprint[2][]{arXiv:#2}

\bibitem[Hamilton et~al\mbox{.}(2017)]%
        {GraphSAGE}
\bibfield{author}{\bibinfo{person}{William~L. Hamilton}, \bibinfo{person}{Zhitao Ying}, {and} \bibinfo{person}{Jure Leskovec}.} \bibinfo{year}{2017}\natexlab{}.
\newblock \showarticletitle{Inductive Representation Learning on Large Graphs}. In \bibinfo{booktitle}{\emph{NeurIPS}}.
\newblock


\bibitem[Jones(1972)]%
        {TF-IDF}
\bibfield{author}{\bibinfo{person}{Karen~Sp{\"{a}}rck Jones}.} \bibinfo{year}{1972}\natexlab{}.
\newblock \showarticletitle{A statistical interpretation of term specificity and its application in retrieval}.
\newblock \bibinfo{journal}{\emph{Journal of Documentation}} \bibinfo{volume}{28}, \bibinfo{number}{1} (\bibinfo{year}{1972}), \bibinfo{pages}{11--21}.
\newblock


\bibitem[Kingma and Ba(2015)]%
        {Adam}
\bibfield{author}{\bibinfo{person}{Diederik~P. Kingma} {and} \bibinfo{person}{Jimmy Ba}.} \bibinfo{year}{2015}\natexlab{}.
\newblock \showarticletitle{Adam: {A} Method for Stochastic Optimization}. In \bibinfo{booktitle}{\emph{ICLR}}.
\newblock


\bibitem[Kipf and Welling(2017)]%
        {GCN}
\bibfield{author}{\bibinfo{person}{Thomas~N. Kipf} {and} \bibinfo{person}{Max Welling}.} \bibinfo{year}{2017}\natexlab{}.
\newblock \showarticletitle{Semi-Supervised Classification with Graph Convolutional Networks}. In \bibinfo{booktitle}{\emph{{ICLR}}}.
\newblock


\bibitem[Li et~al\mbox{.}(2024)]%
        {TQM}
\bibfield{author}{\bibinfo{person}{Haitao Li}, \bibinfo{person}{You Chen}, \bibinfo{person}{Zhekai Ge}, \bibinfo{person}{Qingyao Ai}, \bibinfo{person}{Yiqun Liu}, \bibinfo{person}{Quan Zhou}, {and} \bibinfo{person}{Shuai Huo}.} \bibinfo{year}{2024}\natexlab{}.
\newblock \showarticletitle{Towards an In-Depth Comprehension of Case Relevance for Better Legal Retrieval}. In \bibinfo{booktitle}{\emph{{JSAI}}}, Vol.~\bibinfo{volume}{14741}. \bibinfo{pages}{212--227}.
\newblock


\bibitem[Li et~al\mbox{.}(2023)]%
        {THUIR@COLIEE}
\bibfield{author}{\bibinfo{person}{Haitao Li}, \bibinfo{person}{Weihang Su}, \bibinfo{person}{Changyue Wang}, \bibinfo{person}{Yueyue Wu}, \bibinfo{person}{Qingyao Ai}, {and} \bibinfo{person}{Yiqun Liu}.} \bibinfo{year}{2023}\natexlab{}.
\newblock \showarticletitle{THUIR@COLIEE 2023: Incorporating Structural Knowledge into Pre-trained Language Models for Legal Case Retrieval}.
\newblock \bibinfo{journal}{\emph{CoRR}}  \bibinfo{volume}{abs/2305.06812} (\bibinfo{year}{2023}).
\newblock


\bibitem[Ma et~al\mbox{.}(2023)]%
        {SLR}
\bibfield{author}{\bibinfo{person}{Yixiao Ma}, \bibinfo{person}{Yueyue Wu}, \bibinfo{person}{Qingyao Ai}, \bibinfo{person}{Yiqun Liu}, \bibinfo{person}{Yunqiu Shao}, \bibinfo{person}{Min Zhang}, {and} \bibinfo{person}{Shaoping Ma}.} \bibinfo{year}{2023}\natexlab{}.
\newblock \showarticletitle{Incorporating Structural Information into Legal Case Retrieval}.
\newblock \bibinfo{journal}{\emph{ACM Trans. Inf. Syst.}} (\bibinfo{year}{2023}).
\newblock


\bibitem[Muennighoff et~al\mbox{.}(2022)]%
        {mteb}
\bibfield{author}{\bibinfo{person}{Niklas Muennighoff}, \bibinfo{person}{Nouamane Tazi}, \bibinfo{person}{Lo{\"\i}c Magne}, {and} \bibinfo{person}{Nils Reimers}.} \bibinfo{year}{2022}\natexlab{}.
\newblock \showarticletitle{MTEB: Massive Text Embedding Benchmark}.
\newblock \bibinfo{journal}{\emph{CoRR}}  \bibinfo{volume}{abs/2210.07316} (\bibinfo{year}{2022}).
\newblock


\bibitem[Ponte and Croft(2017)]%
        {LMIR}
\bibfield{author}{\bibinfo{person}{Jay~M. Ponte} {and} \bibinfo{person}{W.~Bruce Croft}.} \bibinfo{year}{2017}\natexlab{}.
\newblock \showarticletitle{A Language Modeling Approach to Information Retrieval}. In \bibinfo{booktitle}{\emph{{SIGIR}}}.
\newblock


\bibitem[Rabelo et~al\mbox{.}(2022a)]%
        {COLIEE}
\bibfield{author}{\bibinfo{person}{Juliano Rabelo}, \bibinfo{person}{Randy Goebel}, \bibinfo{person}{Mi{-}Young Kim}, \bibinfo{person}{Yoshinobu Kano}, \bibinfo{person}{Masaharu Yoshioka}, {and} \bibinfo{person}{Ken Satoh}.} \bibinfo{year}{2022}\natexlab{a}.
\newblock \showarticletitle{Overview and Discussion of the Competition on Legal Information Extraction/Entailment {(COLIEE)} 2021}.
\newblock \bibinfo{journal}{\emph{Rev. Socionetwork Strateg.}} \bibinfo{volume}{16}, \bibinfo{number}{1} (\bibinfo{year}{2022}), \bibinfo{pages}{111--133}.
\newblock


\bibitem[Rabelo et~al\mbox{.}(2022b)]%
        {UA@COLIEE2022}
\bibfield{author}{\bibinfo{person}{Juliano Rabelo}, \bibinfo{person}{Mi{-}Young Kim}, {and} \bibinfo{person}{Randy Goebel}.} \bibinfo{year}{2022}\natexlab{b}.
\newblock \showarticletitle{Semantic-Based Classification of Relevant Case Law}. In \bibinfo{booktitle}{\emph{{JURISIN}}}.
\newblock


\bibitem[Robertson and Walker(1994)]%
        {BM25}
\bibfield{author}{\bibinfo{person}{Stephen~E. Robertson} {and} \bibinfo{person}{Steve Walker}.} \bibinfo{year}{1994}\natexlab{}.
\newblock \showarticletitle{Some Simple Effective Approximations to the 2-Poisson Model for Probabilistic Weighted Retrieval}. In \bibinfo{booktitle}{\emph{{SIGIR}}}.
\newblock


\bibitem[Shao et~al\mbox{.}(2020)]%
        {BERT-PLI}
\bibfield{author}{\bibinfo{person}{Yunqiu Shao}, \bibinfo{person}{Jiaxin Mao}, \bibinfo{person}{Yiqun Liu}, \bibinfo{person}{Weizhi Ma}, \bibinfo{person}{Ken Satoh}, \bibinfo{person}{Min Zhang}, {and} \bibinfo{person}{Shaoping Ma}.} \bibinfo{year}{2020}\natexlab{}.
\newblock \showarticletitle{{BERT-PLI:} Modeling Paragraph-Level Interactions for Legal Case Retrieval}. In \bibinfo{booktitle}{\emph{IJCAI}}.
\newblock


\bibitem[Srivastava et~al\mbox{.}(2014)]%
        {dropout}
\bibfield{author}{\bibinfo{person}{Nitish Srivastava}, \bibinfo{person}{Geoffrey~E. Hinton}, \bibinfo{person}{Alex Krizhevsky}, \bibinfo{person}{Ilya Sutskever}, {and} \bibinfo{person}{Ruslan Salakhutdinov}.} \bibinfo{year}{2014}\natexlab{}.
\newblock \showarticletitle{Dropout: a simple way to prevent neural networks from overfitting}.
\newblock \bibinfo{journal}{\emph{J. Mach. Learn. Res.}} (\bibinfo{year}{2014}).
\newblock


\bibitem[Tang et~al\mbox{.}(2023)]%
        {promptcase}
\bibfield{author}{\bibinfo{person}{Yanran Tang}, \bibinfo{person}{Ruihong Qiu}, {and} \bibinfo{person}{Xue Li}.} \bibinfo{year}{2023}\natexlab{}.
\newblock \showarticletitle{Prompt-based Effective Input Reformulation for Legal Case Retrieval}.
\newblock \bibinfo{journal}{\emph{CoRR}}  \bibinfo{volume}{abs/2309.02962} (\bibinfo{year}{2023}).
\newblock


\bibitem[Tang et~al\mbox{.}(2024a)]%
        {casegnn++}
\bibfield{author}{\bibinfo{person}{Yanran Tang}, \bibinfo{person}{Ruihong Qiu}, \bibinfo{person}{Yilun Liu}, \bibinfo{person}{Xue Li}, {and} \bibinfo{person}{Zi Huang}.} \bibinfo{year}{2024}\natexlab{a}.
\newblock \showarticletitle{CaseGNN++: Graph Contrastive Learning for Legal Case Retrieval with Graph Augmentation}.
\newblock \bibinfo{journal}{\emph{CoRR}}  \bibinfo{volume}{abs/2405.11791} (\bibinfo{year}{2024}).
\newblock


\bibitem[Tang et~al\mbox{.}(2024b)]%
        {casegnn}
\bibfield{author}{\bibinfo{person}{Yanran Tang}, \bibinfo{person}{Ruihong Qiu}, \bibinfo{person}{Yilun Liu}, \bibinfo{person}{Xue Li}, {and} \bibinfo{person}{Zi Huang}.} \bibinfo{year}{2024}\natexlab{b}.
\newblock \showarticletitle{CaseGNN: Graph Neural Networks for Legal Case Retrieval with Text-Attributed Graphs}. In \bibinfo{booktitle}{\emph{ECIR}}.
\newblock


\bibitem[Tang et~al\mbox{.}(2024c)]%
        {caselink}
\bibfield{author}{\bibinfo{person}{Yanran Tang}, \bibinfo{person}{Ruihong Qiu}, \bibinfo{person}{Hongzhi Yin}, \bibinfo{person}{Xue Li}, {and} \bibinfo{person}{Zi Huang}.} \bibinfo{year}{2024}\natexlab{c}.
\newblock \showarticletitle{CaseLink: Inductive Graph Learning for Legal Case Retrieval}. In \bibinfo{booktitle}{\emph{SIGIR}}.
\newblock


\bibitem[van~den Oord et~al\mbox{.}(2018)]%
        {infonce}
\bibfield{author}{\bibinfo{person}{A{\"{a}}ron van~den Oord}, \bibinfo{person}{Yazhe Li}, {and} \bibinfo{person}{Oriol Vinyals}.} \bibinfo{year}{2018}\natexlab{}.
\newblock \showarticletitle{Representation Learning with Contrastive Predictive Coding}.
\newblock \bibinfo{journal}{\emph{CoRR}}  \bibinfo{volume}{abs/1807.03748} (\bibinfo{year}{2018}).
\newblock


\bibitem[Velickovic et~al\mbox{.}(2018)]%
        {GAT}
\bibfield{author}{\bibinfo{person}{Petar Velickovic}, \bibinfo{person}{Guillem Cucurull}, \bibinfo{person}{Arantxa Casanova}, \bibinfo{person}{Adriana Romero}, \bibinfo{person}{Pietro Li{\`{o}}}, {and} \bibinfo{person}{Yoshua Bengio}.} \bibinfo{year}{2018}\natexlab{}.
\newblock \showarticletitle{Graph Attention Networks}. In \bibinfo{booktitle}{\emph{{ICLR}}}.
\newblock


\bibitem[Wang et~al\mbox{.}(2024)]%
        {mistral}
\bibfield{author}{\bibinfo{person}{Liang Wang}, \bibinfo{person}{Nan Yang}, \bibinfo{person}{Xiaolong Huang}, \bibinfo{person}{Linjun Yang}, \bibinfo{person}{Rangan Majumder}, {and} \bibinfo{person}{Furu Wei}.} \bibinfo{year}{2024}\natexlab{}.
\newblock \showarticletitle{Improving Text Embeddings with Large Language Models}. In \bibinfo{booktitle}{\emph{ACL}}.
\newblock


\bibitem[Zhang et~al\mbox{.}(2023)]%
        {cfgl}
\bibfield{author}{\bibinfo{person}{Kun Zhang}, \bibinfo{person}{Chong Chen}, \bibinfo{person}{Yuanzhuo Wang}, \bibinfo{person}{Qi Tian}, {and} \bibinfo{person}{Long Bai}.} \bibinfo{year}{2023}\natexlab{}.
\newblock \showarticletitle{{CFGL-LCR:} {A} Counterfactual Graph Learning Framework for Legal Case Retrieval}. In \bibinfo{booktitle}{\emph{SIGKDD}}.
\newblock


\end{thebibliography}










\end{document}